\documentclass[11pt,twoside]{article}

\usepackage{asp2004}
\usepackage{epsf}
\usepackage{lscape}
\usepackage{graphicx}

\begin{document}
\title{Long-term evolution of accretion disks around the neutron star in Be/X-ray binaries}
 \author{Kimitake~Hayasaki$^{1,2}$ and Atsuo~T.~Okazaki$^3$}
\affil{$^1$Department of Applied Physics, Graduate School of Engineering,
        Hokkaido University, Kitaku N13W8, Sapporo 060-8628, Japan.\\
       $^2$Centre for Astrophysics and Supercomputing, Swinburne University of Technology,
        Hawthorn Victoria 3122 Australia.\\
       $^3$Faculty of Engineering, Hokkai-Gakuen University, Toyohira-ku,
      Sapporo 062-8605, Japan.}

\begin{abstract}
we study the long-term evolution of the accretion disk
around the neutron star in Be/X-ray binaries.
We confirm the earlier result by Hayasaki \& Okazaki (2004) 
that the disk evolves via a two-stage process, which
consists of the initial developing
stage and the later developed stage.
The peak mass-accretion rate is distributed around apastron after the
disk is fully developed.
This indicates that the modulation of the mass accretion rate is
essentially caused by an inward propagation of the one-armed spiral wave.
The X-ray luminosity peak around the apastron could
provide circumstatial evidence for an persistent disk around the
neutron star in Be/X-ray binaries.
\end{abstract}

\keywords{accretion, accretion discs -- hydrodynamics --
          methods: numerical -- binaries: general -- stars: emission-line, 
          Be -- X-rays: binaries}


\vspace{-1.0cm}
\section{Introduction}

Most Be/X-ray binaries show only temporal X-ray activity
, i.e.,
regular periodic outbursts at periastron (Type I) and/or giant
outbursts (Type II) with no orbital modulation.
X-ray outbursts are considered to be caused by the mass-accretion onto
the neutron star through the mass transfer from the Be-star disk (Okazaki et al. 2002).
Recently, Hayasaki \& Okazaki (2004) (hereafter, paper~I) studied the accretion flow
around the neutron star in a Be/X-ray binary with a short period
($P_{\rm{orb}}=24.3~{\rm days}$) and moderate eccentricity
($e=0.34$), using a three-dimensional (3D) 
Smoothed Particle Hydrodynamics (SPH)
code.
They showed that a time-dependent
accretion disk is formed around the neutron star.
Hayasaki \& Okazaki (2005) (hereafter, paper~II) further showed that the disk has a
one-armed spiral structure induced by a phase-dependent mass transfer
from the Be disk.
These are, however, the results from simulations
run over a period shorter than the viscous time-scale of the disk,
and it was not known how the accretion disk evolves over a period 
longer than the viscous time-scale. 
In this paper, we study the long-period evolution 
of accretion disk around the neutron star in Be/X-ray binaries, 
performing the 3D SPH simulations.

\section{Long-term evolution of accretion disk}

Our simulations were performed by using the same 3D SPH code as in
paper~I, which was based on a version originally developed by Benz 
(Benz 1990; Benz et al.\ 1990) and later by Bate, Bonnell \& Price
(1995). 
For the comparison purpose, 
we run the simulations with the same parameters
of model~1 in paper~I, except that the number of injected particles 
per orbit is the three times less than that of model~1 in paper~I.
The orbital period $P_{\rm{orb}}$ is
24.3\,d, the eccentricity $e$ is 0.34, and the Be disc is coplanar
with the orbital plane. 
The inner radius of the simulation region
$r_{\rm{in}}$ is $3.0\times10^{-3}a$, where $a$ is the semi-major axis
of the binary. The polytropic equation of state with the exponent
$\Gamma=1.2$ is adopted. The Shakura-Sunyaev viscosity parameter
$\alpha_{\rm{SS}}=0.1$ throughout the disc.
In what follows, the units of time is $P_{\rm{orb}}$. 

Fig.1 shows the evolution of several non-axisymmetric modes
with an improved definition of the mode strength in paper~II.
The solid-thin, dotted and solid-thick lines denote
the strengths of $m=1$, $m=2$ and $m=3$ modes, respectively.
From Fig.1, we note that the $m=1$ component dominates 
the other components throughout the run.

The top panel of Fig.2 shows the evolution of the mass-accretion rate and 
the corresponding X-ray luminosity, 
whereas the bottom and the middle panel
shows 
orbital-phase dependence of the peak of mass-accretion rate 
and its frequency distribution, respectively.
It is noted from these figures that the mass-accretion rate 
has double peaks per orbit at an initial developing stage ($0\le{t}\la10$). 
While the first peak 
is due to the direct accretion of particles 
with the low specific angular momentum, 
the second one is mainly caused by an inward propagation of $m=1$ mode.
After the disk is fully developed (${t}\ga10$), 
the mass accretion rate 
has a single peak per orbit only due to the wave induced accretion.

\begin{figure}[!ht]
\resizebox{\hsize}{!}
{\includegraphics{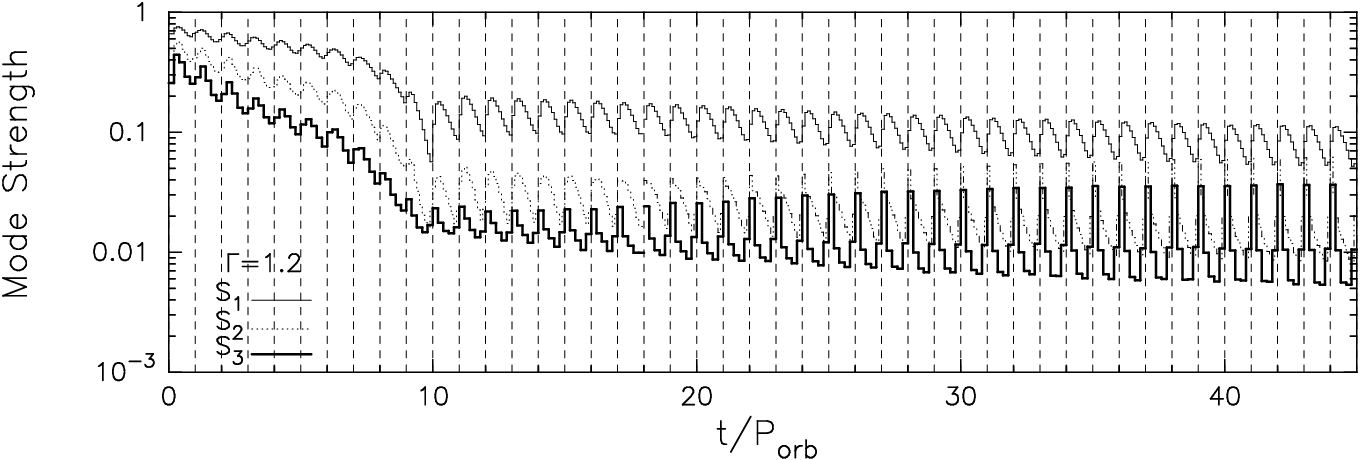}}
\caption{
Evolution of several nonaxisymmetric modes for $0\le{t}\le45$.
The solid-thin, dotted and solid-thick lines denote
the strengths of $m=1$, $m=2$ and $m=3$ modes, respectively.
}
\label{fig1}
\end{figure}


\begin{figure}[!ht]
\begin{center}
\includegraphics*[height=3.4cm]{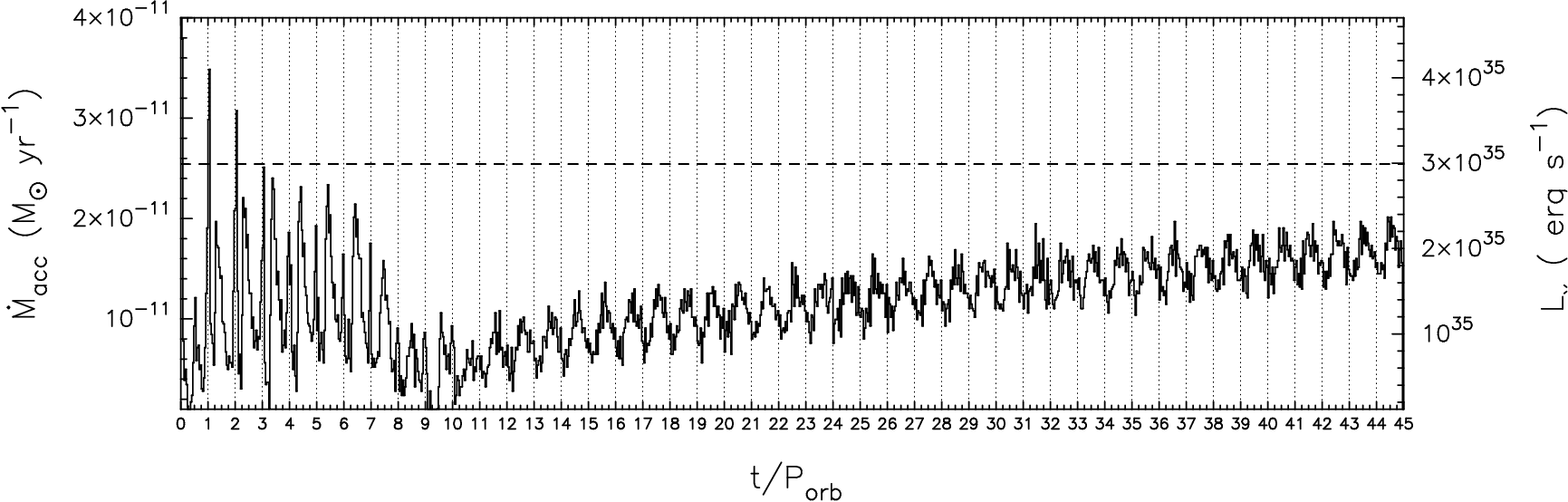}
\includegraphics*[height=4.7cm,angle=0]{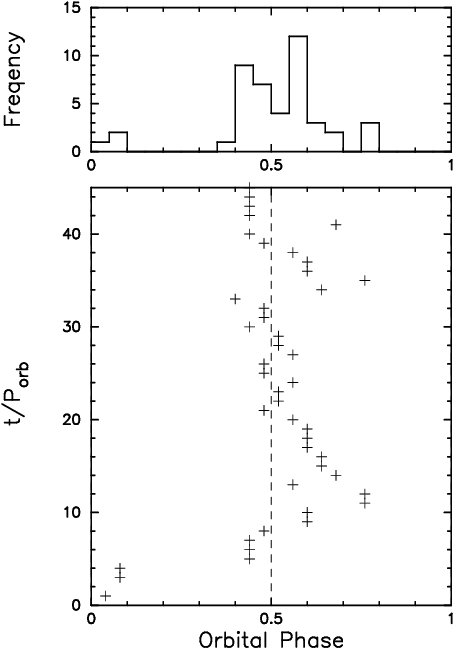}
\end{center}
\caption{
Evolution of the mass-accretion rate $\dot{M}_{acc}$
in units of $M_{\odot}yr^{-1}$ (the top panel) and
orbital-phase dependence of the peak of mass-accretion rate (the bottom panel)
and its frequency distribution (the middle panel).
In the top panel, the right axis is 
for the X-ray luminosity corresponding to 
the mass accretion rate. The horizontal dotted line denotes 
the averaged mass transfer rate from the Be disk.
In the bottom panel, the crosses denote the orbital-phase dependence
 of the X-ray maxima, of which  
the frequency distribution is 
shown by the histogram in the middle panel.
}

\label{fig2}
\end{figure}


\section{Concluding Remarks}

We have performed 3D SPH simulations in order to study
the long-period evolution of the accretion disk
around the neutron star in Be/X-ray binaries.
As shown in left panel of Fig.2, the mass-accretion rate gradually increases 
with the stable material-supply from the Be disk.
This strongly suggests the accretion disk
finally gets to a quasi-steady state as the third evolutionary stage.
On the other hand, the right panel of Fig.2 have shown that
after the disk is fully developed,
the peak of mass-accretion rate is distributed distinct from the periastron
due to an inward propagation of the one-armed spiral wave.
This indicates that 
the X-ray maxima distinct from the periastron could
provide circumstatial evidence for an persistent disk around the
neutron star in Be/X-ray binaries.


\acknowledgements The simulations reported here were performed using the facility 
of the Centre for Astrophysics \& Supercomputing at
Swinburne University of Technology, Australia.
This work has been supported by Grant-in-Aid for the 21st Century
COE Scientific Research Programme on "Topological Science and Technology"
from the Ministry of
Education, Culture, Sport, Science and Technology of Japan (MECSST) and
in part by Grant-in-Aid for Scientific Reserch
(15204010) of Japan Society for the Promotion of Science.



\begin{thebibliography}{}
  \bibitem[Bate, Bonnell \& Price(1995)]{ba}
  Bate M.R., Bonnell I.A.,Price N.M., 1995, MNRAS, 285, 33
  \bibitem[Benz(1990)]{be1}
  Benz W., 1990, in Buchler J. R.,ed.,The Numerical Modelling of Nonlinear Stellar Pulsations.
  Kluwer, Dordrecht, p.269
  \bibitem[Benz et al.(1990)]{be2}
  Benz W., Bowers R.L., Cameron A.G.W., Press W.H., 1990, ApJ, 348, 647
  \bibitem[Hayasaki \& Okazaki(2004)]{haya}
  Hayasaki K \& Okazaki A.T., 2004, MNRAS, 350, 971
  \bibitem[Hayasaki \& Okazaki(2005)]{haya2}
  Hayasaki K \& Okazaki A.T., 2005, MNRAS, 360L, 15
  \bibitem[Okazaki et al.(2002)]{oka2}
  Okazaki A.T., Bate M.R., Ogilvie G.I \& Pringle J.E., 2002, MNRAS, 337, 967
\end{thebibliography}
\end{document}